\documentclass{iau}
\usepackage{graphicx}

\title[Polarization in Mira variable stars] 
{Long-term polarization observations of Mira variable stars suggest asymmetric structures}

\author[Neilson, Ignace, \& Henson]   
{Hilding R.~Neilson,
 Richard Ignace \and Gary D.~Henson}

\affiliation{Dept.~of Physics \& Astronomy, East Tennessee State University, PO Box 70300, Johnson City, TN 37614, USA \\ email: {\tt neilsonh@etsu.edu}}

\pubyear{2013}
\volume{301}  
\pagerange{???--???}
\jname{Precision Asteroseismology: Celebration of the Scientific Opus of Wojtek Dziembowski}
\editors{W.~Chaplin, J.~Guzik, G.~Hander \& A.~Pigulski, eds.}
\begin{document}

\maketitle
\begin{abstract}
Mira and semi-regular variable stars have been studied for centuries but continue to be enigmatic. One unsolved mystery is the presence of polarization from these stars. In particular, we present 40 years of polarization measurements for the prototype o~Ceti and V~CVn and find very different phenomena for each star. The polarization fraction and position angle for Mira is found to be small and highly variable. On the other hand, the polarization fraction for V~CVn is large and variable, from 2 - 7 \%, and its position angle is approximately constant, suggesting a long-term asymmetric structure. We suggest a number of potential scenarios to explain these observations.
\end{abstract}

\firstsection 
\section{Introduction}
Mira and semi-regular variable stars have been observed almost continuously for centuries since the discovery by Fabricius in 1596 that the prototype was variable; although there is some evidence that variability was detected by earlier civilizations (e.g., \cite[Sahade \& Wood, 1978]{Sahade1978}).

These continuous observations have been important for the understanding of stellar pulsation and shedding insight into the theory of stellar evolution (\cite[Uttenthaler et al. 2011]{Uttenthaler2011}). Furthermore, the continuous observations have detected irregular pulsation in many evolved red giant stars. Mira variable stars are also important standard candles, that can be employed to measure distances to other galaxies (\cite[Whitelock et al. 2008]{Whitelock2008}).

Another probe to explore the structure and evolution of Mira variable stars is using linear polarization observations to probe asymmetric stellar properties (\cite[Serkowski \& Shawl 2001]{Serkowski2001}).

\section{Polarization Data}
We present polarization and position angle observations of the prototype Mira and the semi-regular variable star V~CVn. These quantities are derived from observed Stokes $Q$ and $U$ parameters, where the polarization $p = (Q^2+U^2)^{1/2}$,and the position angle $\psi = \tan^{-1}(U/Q)$. Using these parameters we explore asymmetric stellar structures, i.e., deviations from circular symmetry of the radiation field.
\begin{figure}[t]
\begin{center}
 \includegraphics[width=0.49\textwidth]{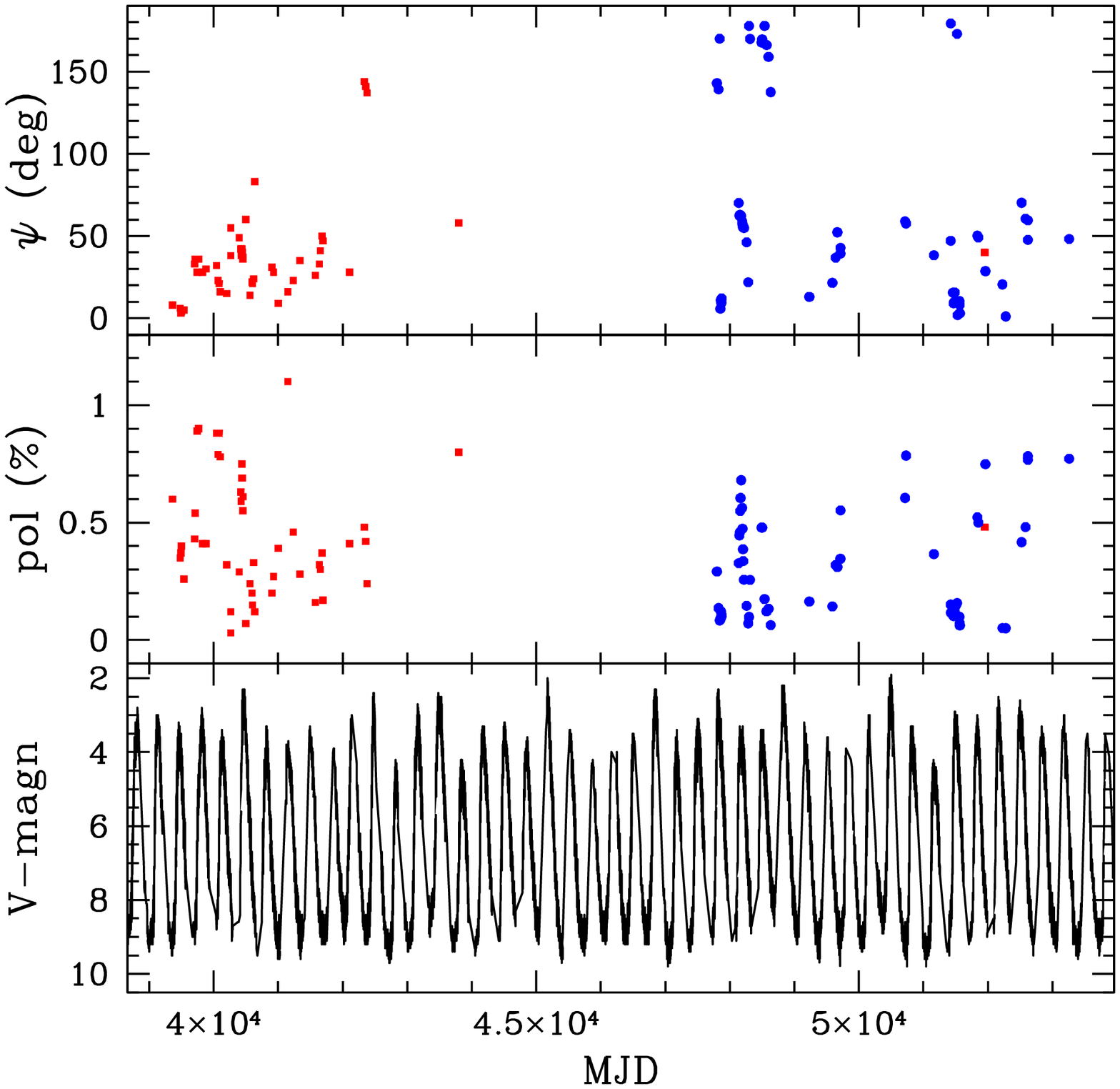} \includegraphics[width=0.49\textwidth]{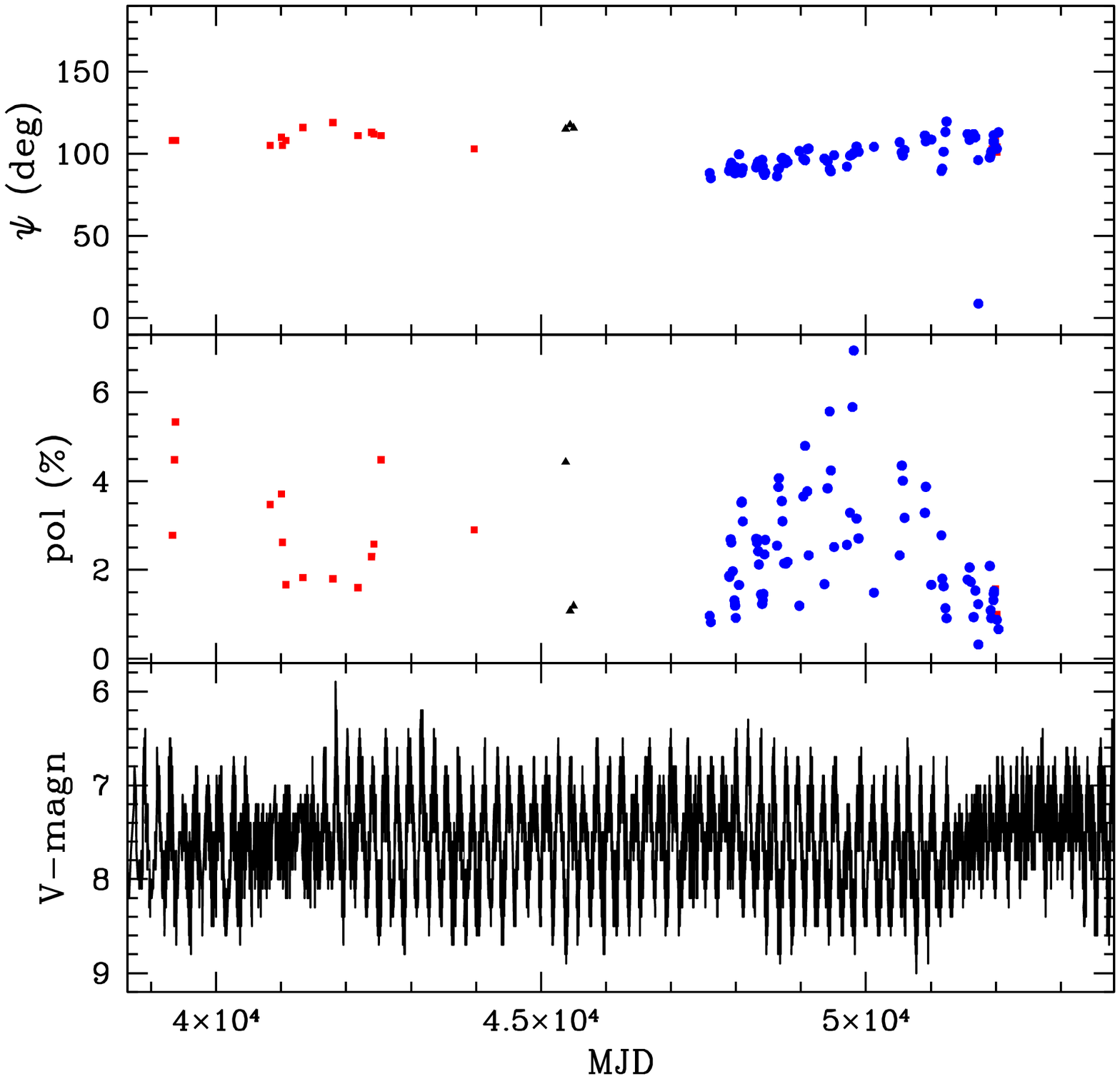} 
 \caption{Measured position angle, $\psi$, in degrees; polarization percentage; and visual brightness for Mira (Left) and V~CVn (Right) from the AAVSO database. Circles represent HPOL observations, triangles are those from \cite[Magalh\~aes et al. (1986a)]{Magalhaes1986a} and the squares from \cite[Poliakova (1981)]{Poliakova1981}.}
   \label{fig1}
\end{center}
\end{figure}

We plot in Fig.~1 the measured values of $p$, $\psi$, and V-band brightness variations for Mira (left) and V~CVn (right) spanning a time scale of forty years. The latter data, denoted by dots is taken from the HPOL database (\cite[Wolff et al. 1996]{Wolff1996}) while the triangles are from \cite[Magalh\~aes et al. (1986a)]{Magalhaes1986a} and the squares from \cite[Poliakova (1981)]{Poliakova1981}. The V-band observations are from the AAVSO database. For Mira, $p$ varies from $0 $ -- $ 1\%$ with large variation of $\psi$, whereas, $p = 2$ -- $7\%$ for
V~CVn with a nearly constant value of $\psi$ spanning 40 years.

\section{Future Work}
We suggest that the polarization and position angle measurements for V~CVn is consistent with the structure of a circumstellar disk (Neilson et al. submitted), whereas the polarization for Mira is due to convective hot spots (Ignace et al. in prep).

We are also exploring polarization measurements of other Mira and semi-regular variable stars, such as R Leo. There are about 20 such stars in the HPOL database. However, none of the other stars in the HPOL database have detect polarization at similar levels as
V~CVn. Only the star L$_2$ Pup appears to have similar polarization (\cite[Magalh\~aes et al. 1986b]{Magalhaes1986b}), suggesting that circumstellar disks about these stars is rare but is consistent with observations of disks about white dwarf stars (\cite[Farihi et al. 2005]{Farihi2005}).

We continue to observe V~CVn, Mira and a sample of variables in the northern hemisphere using the Lulin One-Meter Telescope in Taiwan.  We are grateful for funding from NSF Grant AST-0807664 and to the many observers who contribute data to the AAVSO.

\end{document}